\documentclass[conference]{IEEEtran}
\usepackage{graphicx}
\usepackage{cite}
\usepackage{amsmath}
\usepackage{amsfonts}
\usepackage{amssymb}
\usepackage{psfrag}
\usepackage{color}
\usepackage{multicol}
\usepackage{amsthm}

\usepackage{tikz}

\newcommand\norm[1]{\left\lVert#1\right\rVert}		
\newcommand{\bm}{\mathbf}
\newcommand{\be}{\begin{equation}}
\newcommand{\ee}{\end{equation}}
\newcommand{\bse}{\begin{subequations}}
\newcommand{\ese}{\end{subequations}}
\newcommand{\bea}{\begin{eqnarray}}
\newcommand{\eea}{\end{eqnarray}}
\newcommand{\x}{{\bm x}}

\newcommand{\ba}{{\bm a}}
\newcommand{\bb}{{\bm b}}
\newcommand{\bc}{{\bm c}}
\newcommand{\br}{{\bm r}}

\newcommand{\dd}{{\bm d}}
\newcommand{\bA}{{\bm A}}

\newcommand{\bF}{{\bf F}}
\newcommand{\bD}{{\bf D}}

\newcommand{\bG}{{\bf G}}
\newcommand{\bH}{{\bf H}}

\newcommand{\bg}{{\bf g}}

\newcommand{\h}{{\bf h}}
\newcommand{\bh}{{\bf h}}

\newcommand{\bd}{{\bf d}}

\newcommand{\bff}{{\bf f}}
\newcommand{\bzero}{{\bf 0}}

\newcommand{\I}{{\bm I }}

\newcommand{\BG}{{\boldsymbol{\mathcal G}}}

\newcommand{\bpsi}{\mbox{\boldmath$\psi$}}
\newcommand{\bnu}{\mbox{\boldmath$\nu$}}
\newcommand{\bgamma}{\mbox{\boldmath$\gamma$}}

\newcommand{\bxi}{\mbox{\boldmath{$\xi$}}}

\newcommand{\brho}{\mbox{\boldmath{$\rho$}}}
\theoremstyle{definition}

\begin{document}
\title{Time Reversal with Post-Equalization for OFDM without CP in Massive MIMO}

\author{\normalsize  Arman Farhang$^*$, Amir Aminjavaheri$^\dagger$, Ahmad RezazadehReyhani$^\dagger$, Linda E. Doyle$^*$ and Behrouz Farhang-Boroujeny$^\dagger$  
\\$^*$CONNECT, Trinity College Dublin, Ireland, \\
$^\dagger$ECE Department, University of Utah, USA. \\
Email: \{farhanga, ledoyle\}@tcd.ie, \{aminjav, rezazade, farhang\}@ece.utah.edu }

\maketitle

\begin{abstract}
This paper studies the possibility of eliminating the redundant cyclic prefix (CP) of orthogonal frequency division multiplexing (OFDM) in massive multiple-input multiple-output systems. The absence of CP increases the bandwidth efficiency in expense of intersymbol interference (ISI) and intercarrier interference (ICI). It is known that in massive MIMO, different types of interference fade away as the number of base station (BS) antennas tends to infinity. In this paper, we investigate if the channel distortions in the absence of CP are averaged out in the large antenna regime. To this end, we analytically study the performance of the conventional maximum ratio combining (MRC) and realize that there always remains some residual interference leading to saturation of signal to interference (SIR). This saturation of SIR is quantified through mathematical equations. Moreover, to resolve the saturation problem, we propose a technique based on time-reversal MRC with zero forcing multiuser detection (TR-ZF). Thus, the SIR of our proposed TR-ZF does not saturate and is a linear function of the number of BS antennas. We also show that TR-ZF only needs one OFDM demodulator per user irrespective of the number of BS antennas; reducing the BS signal processing complexity significantly. Finally, we corroborate our claims as well as analytical results through simulations.
\end{abstract}

\section{Introduction}\label{sec:Introduction}
Massive multiple-input multiple-output (MIMO) is a multiuser technique that has recently emerged as a strong candidate technology for the fifth generation of wireless networks (5G), \cite{Marzetta2010,Rusek2013}. One of the key features of massive MIMO is that the effects of noise and multiuser interference (MUI) fade away as the number of base station (BS) antennas tends to infinity, \cite{Marzetta2010}. Furthermore, if the number of BS antennas $M$ is much larger than the number of users $K$, optimal performance can be achieved through the most straightforward detection/precoding techniques, \cite{Marzetta2010}. Another appealing property of massive MIMO is that it allows the users to simultaneously utilize the same resources in time and frequency. This brings a substantial improvement in the capacity of multiuser networks.

Due to the law of large numbers, different types of interference such as thermal noise, MUI, and hardware imperfections average out as a result of combining the received signals from a large number of antennas at the base station \cite{Rusek2013,Larsson2014,Bjornson2015}. To convert the frequency selective channels between each mobile terminal (MT) antenna and BS antennas into a set of flat fading channels over each subcarrier band, orthogonal frequency division multiplexing (OFDM) with cyclic prefix (CP) has been extensively utilized in massive MIMO literature, e.g. \cite{Marzetta2010,Xiang2015}. Therefore, the MT data streams can be distinguished from each other through the respective channel responses between their antennas and the BS antennas.

In OFDM, the symbol duration $T$ includes CP which adds an extra overhead to the system. Hence, removing CP can improve the overall spectral efficiency of the network.
As removing CP reduces the length of each packet, hence, in a time-division duplex (TDD) mode, which is usually assumed in the massive MIMO literature \cite{Marzetta2010}, it reduces the impact of the channel aging as the network switches between the uplink and downlink transmission modes. Therefore, in this context, it is highly desirable to eliminate the CP from OFDM.
 Eliminating CP comes in expense of intersymbol interference (ISI) as well as intercarrier interference (ICI) that are imposed by the transient of the multipath channel. To tackle the ISI and ICI problem of OFDM in the absence of CP or with insufficient CP, a number of interference cancellation based techniques exist in the literature, \cite{Molisch2007,Beheshti2009}. However, such approaches become prohibitively complex when a large number of antennas are utilized at the BS.

In the recent past, a number of attempts have been made to shorten the CP length and hence improve bandwidth efficiency of OFDM through utilization of time-reversal (TR) precoding/detection in multiple input single output (MISO) and single input multiple output (SIMO) systems, \cite{Gomes2008,Zhiqiang2012,Dubois2013,Maaz2015}. TR is a technique based on the principle of channel reciprocity that is extensively used in the underwater acoustic channels, \cite{Gomes2008}. However, application of TR is recent to wireless channels, especially massive MIMO, \cite{Dubois2013,Maaz2015,Pitarokoilis2015,Pitarokoilis2012}. 
In \cite{Zhiqiang2012}, the authors propose a CP length design method to satisfy certain performance requirements in underwater acoustic channels. Their method balances the tradeoff between the CP length and the injected interference due to the the residual ISI and ICI imposed by the insufficient CP. As it is shown in \cite{Zhiqiang2012}, time reversed OFDM (TR-OFDM) has a reduced implementation complexity compared with the conventional implementations as only one OFDM demodulator is required. However, in \cite{Zhiqiang2012}, the authors do not look into the multiuser scenario. In addition, they do not study TR-OFDM in the context of massive MIMO as the largest number of BS antennas they consider is eight. In contrast, the authors of \cite{Dubois2013} study TR precoding in a large MISO-OFDM system where they show the CP length can be significantly reduced, thanks to the time localization property of TR. In \cite{Dubois2013}, the TR precoding is performed in the frequency domain. Thus, an OFDM demodulator is utilized per antenna branch. This imposes a substantial amount of computational complexity when a large number of BS antennas are deployed, i.e., in the order of hundreds. In a more recent work, the authors of \cite{Maaz2015} analytically studied the bit error rate (BER) performance for the TR precoding method of \cite{Dubois2013}. According to the above discussion, even though there has been a number of attempts to shorten the CP and consequently improve the spectral efficiency of OFDM, to the best of our knowledge, there is no study investigating the possibility to completely eliminate the CP in massive MIMO. Therefore, the major contribution of this paper is to show that it is possible to successfully remove the redundant CP of OFDM in massive MIMO as the channel distortions fade away in the large antenna regime.

Recently, the authors in \cite{Farhang2014} have raised the notion of self-equalization in filter bank multicarrier (FBMC) under massive MIMO channels, i.e., linear combining of the received signals at different antennas averages out the channel distortions. 
Pursuing the same analogy as that of \cite{Farhang2014}, in this paper, we are trying to investigate if massive MIMO can average out the distortions due to the multipath channel, i.e., ISI and ICI, in OFDM without CP. To this end, in this paper, we mathematically analyze the asymptotic signal to interference ratio (SIR) of the conventional equalization for OFDM in the absence of CP as the number of BS antennas tends to infinity, i.e., maximum ratio combining (MRC) of the received signals at different BS antennas through single-tap equalization per subcarrier, \cite{Marzetta2010}. As a result, we derive an asymptotic closed-form SIR relationship for the conventional equalization as a function of the power delay profile of the channel. We show that when the conventional MRC is applied, the channel distortions due to the absence of CP do not completely average out and there exists residual ISI and ICI even with infinite number of BS antennas. Hence, a key finding of this paper is that through MRC, SIR saturates at a certain level. To resolve the saturation problem, we propose an algorithm based on time-reversal MRC (TR-MRC) combined with a zero forcing based post-equalization/multiuser detection which we call TR-ZF. In addition, we present the baseband implementation of our proposed algorithm where we show only one OFDM demodulator is required per user irrespective of the number of BS antennas. This significantly simplifies the BS structure compared to the existing structures in the literature, e.g., \cite{Dubois2013}. It is worth mentioning that the same procedure as in TR-ZF can be applied to the downlink where we have two stages of precoding, 1) ZF precoding, 2) TR maximum ratio transmission (TR-MRT). Finally, we evaluate the performance of our proposed technique through numerical simulations and compare it with the conventional equalization techniques. Based on our simulations, the signal to interference plus noise (SINR) does not saturate when the proposed TR-ZF technique is used. The results show that SINR of our proposed technique is a linear function of the number of BS antennas. We also numerically study the bit error rate (BER) performance of our proposed technique together with the conventional MRC and compare them with a benchmark that is CP-OFDM. CP-OFDM is chosen as a benchmark since its SIR is infinity due to the presence of a CP longer than the channel delay spread. Our BER results show that while the conventional MRC has a poor performance, the propose TR-ZF has a performance very close to that of CP-OFDM.

The rest of the paper is organized as follows. Section~\ref{sec:System_Model} presents the system model for OFDM without CP in massive MIMO. Section~\ref{sec:SIR_Analysis} includes the asymptotic SIR analysis of the conventional equalization. Section~\ref{sec:TRZF}, explains our proposed TR-ZF algorithm. The efficacy of our proposed technique is analyzed in Section~\ref{sec:Numerical_Results} through computer simulations. Finally, the conclusions are drawn in Section~\ref{sec:Conclusions}.

\textit{Notations:} Matrices, vectors and scalar quantities are denoted by boldface uppercase, boldface lowercase and normal letters, respectively. $[\bA]_{m,n}$ represents the element in the $m^{\rm{th}}$ row and $n^{\rm{th}}$ column of $\bA$ and $\bA^{-1}$ signifies the inverse of $\bA$. $\I_M$ and ${\bf{0}}_{M\times N}$ are the identity and zero matrices of the sizes $M\times M$ and $M\times N$, respectively. $\bD={\rm diag}(\ba)$ is a diagonal matrix whose diagonal elements are formed by the elements of the vector $\ba$. The superscripts $(\cdot)^{\rm T}$, $(\cdot)^{\rm H}$ and $(\cdot)^\ast$ indicate transpose, conjugate transpose and conjugate operations, respectively. Finally, $|\cdot|$ is the absolute value operator on scalers and the element-wise absolute value operator when applied to the vectors, and $\ast$ represents linear convolution.

\section{System Model}\label{sec:System_Model}
We consider a large scale MIMO system similar to the one discussed in \cite{Marzetta2010}. In the scenario of interest to this paper, $K$ MTs are simultaneously communicating with the BS under the time division duplexing (TDD) regime. Each MT has a single transmit and receive antenna while the BS is equipped with $M\gg K$ transmit/receive antennas. In this paper, we assume OFDM modulation for data transmission with the total number of $N$ subcarriers. To increase the bandwidth efficiency, opposed to \cite{Marzetta2010}, we do not append CP in the beginning of the OFDM symbols.

Let the vectors $\x_k = [\x_{k,0}^{\rm T},\ldots,\x_{k,Q-1}^{\rm T}]^{\rm T}$ include $Q$ concatenated OFDM symbols where $\x_{k,i}=\bF_N^{\rm H}\bd_{k,i}$ is OFDM modulated symbol $i$ belonging to user $k$, and $\bF_N$ is the $N$-point normalized discrete Fourier transform (DFT) matrix. $\bd_{k,i}=[d_{k,i}(0),\ldots,d_{k,i}(N-1)]^{\rm T}$ is the transmit data vector of user $k$ on symbol $i$ whose elements are independent and identically distributed (i.i.d.) complex random variables with the variance of unity chosen from a quadrature amplitude modulation (QAM) constellation. The received signal at the $m^{\rm th}$ BS antenna after going through the multipath channel can be obtained as 
\be\label{eqn:rm}
\br_m = \sum^{K-1}_{k=0}\x_k\ast \h_{k,m}+\bnu_m,
\ee
where the vector $\h_{k,m}$ includes the samples of the channel impulse response (CIR) between user $k$ and BS antenna $m$ whose length is $L$. Independent frequency selective fading channels are assumed between each MT antenna and the BS antennas. Hence, each MT is distinguished by the BS using the channel responses between its antenna and the BS antennas.
$\bnu_m$ is the complex additive white Gaussian noise (AWGN) vector at the $m^{\rm th}$ BS antenna, i.e., $\boldsymbol\nu_m~{\sim}~{\mathcal {CN}}(0,{{\sigma_\nu}^{2}}{\I_{P}})$, ${\sigma_\nu}^{2}$ is the noise variance at the input of the BS antennas, and $P=N	Q+L-1$ is the length of the received signal $\br_m$. 

The channel impulse responses, $\h_{k,m}$, can be modeled as linear time-invariant finite impulse response filters with the order of $L$, i.e., $\h_{k,m}=[h_{k,m}(0),\ldots,h_{k,m}(L-1)]^{\rm T}$ where the elements $h_{k,m}(\ell)$ are i.i.d. complex Gaussian random variables with zero mean and variance of $\rho(\ell)$. Thus, $\h_{k,m}~{\sim}~{\mathcal {CN}}(0,{\rm diag}({\brho}))$ where the vector $\brho=[\rho(0),\ldots,\rho(L-1)]^{\rm T}$ is the power delay profile (PDP) of the channel model. Throughout the paper, we assume a normalized PDP, i.e., $\sum_{\ell=0}^{L-1}\rho(\ell)=1$. 
{\begin{figure*}
\vspace{-0.7 cm}
\begin{multicols}{2}
{
\scriptsize
\be
 \bH_{k,m}^{\rm ISI} =
 \begin{pmatrix}
  0 & \cdots & h_{k,m}(L-1) & h_{k,m}(L-2) & \cdots & h_{k,m}(1) \\
	0 & \cdots & 0   & h_{k,m}(L-1) & \cdots & h_{k,m}(2)  \\
  \vdots & \ddots  & \vdots & \vdots & \ddots & \vdots \\
  0      & \cdots  &  \cdots     & \cdots & \cdots & h_m(L-1)\\
	0      & \cdots & \cdots  &  \cdots     & \cdots & 0\\
	\vdots  & \vdots  & \vdots & \vdots & \ddots & \vdots \\
	0 & \cdots & \cdots  &  \cdots & \cdots & 0 
 \end{pmatrix},\nonumber
\ee
}
{
\scriptsize
\be
\vspace{-0.6 cm}
 \bH_{k,m}^{\rm ICI} =
 \begin{pmatrix}
  h_{k,m}(0) & 0   & 0 & \cdots & 0 \\
	h_{k,m}(1) & h_{k,m}(0) & 0 & \cdots & 0 \\
  \vdots  & \vdots  & \vdots & \ddots & \vdots \\
  h_{k,m}(L-1) & h_{k,m}(L-2) & h_{k,m}(L-3) & \cdots & 0 \\
	0 & h_{k,m}(L-1) & h_{k,m}(L-2) & \cdots & 0 \\
	\vdots  & \vdots  & \vdots & \ddots & \vdots \\
	0 & 0 & 0 & \cdots & h_{k,m}(0) 
 \end{pmatrix}.\nonumber
\ee
}
\end{multicols}
\hrulefill
\end{figure*}
}

Considering perfect timing and frequency synchronization, the received signal at the $m^{\rm th}$ BS antenna within a rectangular time window of size $N$ samples at the position of symbol $i$ can be written as
\be\label{eqn:rmi}
\br_{m,i} = \sum^{K-1}_{k=0} \bH_{k,m}^{\rm ISI}\x_{k,i-1}+\bH_{k,m}^{\rm ICI}\x_{k,i}+{\bnu}_{m,i},
\ee
where the $N\times N$ toeplitz matrices $\bH_{k,m}^{\rm ISI}$ and $\bH_{k,m}^{\rm ICI}$ for a given user, $k$, when multiplied to the vectors $\x_{k,i-1}$ and $\x_{k,i}$, create the tail of the symbol $i-1$ overlapping with $L-1$ samples in the beginning of the $i^{\rm th}$ symbol and the channel affected symbol $i$, respectively. $\bH_{k,m}^{\rm ISI}$ and $\bH_{k,m}^{\rm ICI}$ are time domain ISI and ICI matrices, respectively, with the structures shown on the top of this page. The vector ${\bnu}_{m,i}$ includes $N$ samples of the AWGN vector $\bnu_m$ at the position of symbol $i$.

\section{SIR Analysis of Conventional Equalization}\label{sec:SIR_Analysis}
In this section, we derive an analytical SIR formula for the conventional OFDM equalization and study the asymptotic SIR behaviour as the the number of BS antennas tends to infinity. Without loss of generality and for the sake of simplicity, we consider only one active user. Thus, we drop the user indices and rearrange (\ref{eqn:rmi}) as
 \be\label{eqn:rmiu1}
\br_{m,i} =  \bH_{m}^{\rm ISI}\x_{i-1}+\bH_{m}^{\rm ICI}\x_{i}+{\bnu}_{m,i}.
\ee

In conventional equalization, the received signals at the BS antennas are passed through OFDM demodulators, one per antenna, and then the output of the DFT bins are combined using DFT coefficients of the CIRs between the MT and BS antennas. To cast this procedure into a mathematical formulation and pave the way for our SIR derivations, we start from equation (\ref{eqn:rmiu1}) and obtain the output of the OFDM demodulator at antenna $m$ as
\bea\label{eqn:rmibar}
\tilde{\br}_{m,i} &=& \bF_N\bH_{m}^{\rm ISI}\x_{i-1}+\bF_N\bH_{m}^{\rm ICI}\x_i+\bF_N{\bnu}_{m,i}\nonumber\\
&=& \bF_N\bH_{m}^{\rm ISI} \bF_N^{\rm H}\bd_{i-1}+\bF_N\bH_{m}^{\rm ICI} \bF_N^{\rm H}\bd_i+\bF_N{\bnu}_{m,i}\nonumber\\
&=&\widetilde{\bH}_{m}^{\rm ISI} \bd_{i-1}+\widetilde{\bH}_{m}^{\rm ICI} \bd_i+\tilde{\bnu}_{m,i},
\eea
where $\widetilde{\bH}^{\rm ISI}_{m}= \bF_N\bH_{m}^{\rm ISI} \bF_N^{\rm H}$ and $\widetilde{\bH}^{\rm ICI}_{m}=\bF_N\bH_{m}^{\rm ICI} \bF_N$ are $N\times N$ frequency domain ISI and ICI matrices at the $m^{\rm th}$ BS antenna, respectively. $\tilde{\bnu}_{m,i}=\bF_N{\bnu}_{m,i}$ is the AWGN vector for symbol $i$ in the frequency domain and $\tilde{\br}_{m,i}=[\tilde{r}_{m,i}(0),\ldots,\tilde{r}_{m,i}(N-1)]^{\rm T}$. The elements of $\widetilde{\bH}^{\rm ICI}_{m}$ and $\widetilde{\bH}^{\rm ISI}_{m}$ can be calculated as
\be\label{eqn:H_ICI}
[\widetilde{\bH}^{\rm ICI}_{m}]_{pq} = \frac{1}{N}\sum^{N-1}_{n=0}\sum^{L-1}_{\ell=0}h_m(\ell)e^{j\frac{2\pi}{N}(nq-\ell q-np)}w(n-\ell),
\ee
and
\be\label{eqn:H_ISI}
[\widetilde{\bH}^{\rm ISI}_{m}]_{pq} = \frac{1}{N}\sum^{N-1}_{n=0}\sum^{L-1}_{\ell=0}h_m(\ell)e^{j\frac{2\pi}{N}(nq-\ell q-np)}w(n-\ell+N),
\ee
where $w(n)$ is the window function. Here, we assume
\be\label{eqn:window}
w(n)=
\left\{
	\begin{array}{ll}
	1, &0\leqslant n \leqslant N-1, \\
	0, & \rm{otherwise}.
	\end{array}
\right. 
\ee

Let the vector $\tilde{\bh}_m = \bF_N\bar{\bh}_m=[\tilde{h}_m(0),\ldots,\tilde{h}_m(N-1)]^{\rm T}$ contain the $N$-point DFT coefficients of the CIR between the MT and BS antenna $m$ where the $N\times 1$ vector $\bar{\bh}_m$ is the zero-padded version of ${\bh}_m$. The outputs of a given DFT bin, $p$, in OFDM demodulators at different BS antennas, i.e. $\tilde{r}_{m,i}(p)$ for $m=0,\ldots,M-1$, are combined using the combining vector 
\be\label{eqn:MF_MRC}
\bpsi_p = \frac{\bgamma_p}{\norm{\bgamma_p}^2},
\ee
where $\bgamma_p = [\tilde{h}_0(p),\ldots,\tilde{h}_{M-1}(p)]^{\rm T}$. Hence, the combined output at frequency bin $p$ can be obtained as
\be\label{eqn:MRC_p}
\hat{d}_i(p) = \bpsi_p^{\rm H}\tilde{\br}_{i}(p),
\ee
where $\tilde{\br}_{i}(p) = [\tilde{r}_{0,i}(p),\ldots,\tilde{r}_{M-1,i}(p)]^{\rm T}$. 

To study the effect of ICI and ISI on the output of the maximum-ratio combiner (\ref{eqn:MRC_p}), in the following subsections, we first derive the signal, ICI and ISI powers and then present a closed-form relationship for the asymptotic SIR as the number of BS antennas tends to infinity.

\subsection{Signal power calculation}\label{subsec:Ps_MRC}
Using (\ref{eqn:MRC_p}), (\ref{eqn:MF_MRC}) and (\ref{eqn:rmibar}), the signal power in a given subcarrier $p$ after combining can be obtained as
\be\label{eqn:Ps_MRC}
P_{\rm s} = |\bpsi_p^{\rm H}\bxi|^2~~\xrightarrow{M\rightarrow\infty}~~|\mathbb{E}\{\tilde{h}_m^\ast (p)[\widetilde{\bH}^{\rm ICI}_{m}]_{pp}\}|^2,
\ee
where $\bxi = [[\widetilde{\bH}^{\rm ICI}_{0}]_{pp},\ldots,[\widetilde{\bH}^{\rm ICI}_{M-1}]_{pp}]^{\rm T}$ and the right hand side of equation (\ref{eqn:Ps_MRC}) is obtained through the law of large numbers. Substituting $[\widetilde{\bH}^{\rm ICI}_{m}]_{pp}$ from (\ref{eqn:H_ICI}) in (\ref{eqn:Ps_MRC}), $P_{\rm s}$ can be expanded as
\bea\label{eqn:Ps_expectation}
P_{\rm s} &=& |\mathbb{E}\{\bar{h}_m^\ast (p)[\widetilde{\bH}^{\rm ICI}_{m}]_{pp}\}|^2  \nonumber \\
&=& \bigg|\frac{1}{N}\mathbb{E}\bigg\{\sum^{N-1}_{n=0}\sum^{L-1}_{\ell=0}\tilde{h}_m^\ast (p)h_m(\ell)e^{-j\frac{2\pi}{N}\ell p}w(n-\ell)\bigg\}\bigg|^2\nonumber \\
&=& \bigg|\frac{1}{N}\mathbb{E}\bigg\{\sum^{N-1}_{n=0}\sum^{L-1}_{\ell=0}\sum^{L-1}_{\ell^{\prime}=0}{h}_m^\ast (\ell^{\prime})h_m(\ell)e^{-j\frac{2\pi}{N}(\ell-\ell^{\prime}) p} \nonumber \\
&&\times w(n-\ell)\bigg\}\bigg|^2 \nonumber \\
&=&\bigg|\frac{1}{N}\sum^{L-1}_{\ell=0}(N-\ell)\rho(\ell)\bigg|^2=(1-\frac{\bar{\tau}}{N})^2,
\eea
where $\bar{\tau} = \sum^{L-1}_{\ell=0}\ell \rho(\ell) $
is the average delay spread of the channel. From equation (\ref{eqn:Ps_expectation}), one may realize that $P_{\rm s}$ is the same for all the subcarriers and it only depends on the average delay spread of the channel.

\subsection{ICI power calculation}\label{subsec:PICI_MRC}
In this subsection, we derive the ICI power from a given subcarrier, $q$, on subcarrier $p$ after combining. Defining the subcarrier distance as $d \triangleq\{(q-p)\mod N\}$, asymptotic ICI power from subcarrier $q$ with the modulo-$N$ distance $d$ from subcarrier $p$ as the number of BS antennas tends to infinity can be obtained as
\be\label{eqn:PICI_MRC}
P_{\rm ICI}(d) ~~\xrightarrow{M\rightarrow\infty}~~ |\mathbb{E}\{\tilde{h}_m^\ast (p)[\widetilde{\bH}^{\rm ICI}_{m}]_{pq}\}|^2.
\ee
In order to derive a closed-form for equation (\ref{eqn:PICI_MRC}), we expand $\zeta = \mathbb{E}\{\tilde{h}_m^\ast (p)[\widetilde{\bH}^{\rm ICI}_{m}]_{pq}\}$ using (\ref{eqn:H_ICI}) in the following.
\bea\label{eqn:PICI_MRC1}
\zeta&=&\frac{1}{N}\mathbb{E}\bigg\{\sum^{N-1}_{n=0}\sum^{L-1}_{\ell=0}\sum^{L-1}_{\ell^{\prime}=0}{h}_m^\ast (\ell^{\prime})h_m(\ell)e^{-j\frac{2\pi}{N}(\ell q - nq - \ell^{\prime}p + np)}\nonumber \\
&& \times w(n-\ell)\bigg\} \nonumber \\
&=&\frac{1}{N}\sum^{N-1}_{n=0}\sum^{L-1}_{\ell=0}\rho(\ell)e^{-j\frac{2\pi}{N}(\ell-n)d}w(n-\ell) \nonumber \\
&=&\frac{1}{N}\sum^{L-1}_{\ell=0}\rho(\ell)e^{-j\frac{2\pi\ell d}{N}}\sum^{N-1}_{n=\ell}e^{-j\frac{2\pi nd}{N}}\nonumber \\
&=&-\frac{1}{N}\sum^{L-1}_{\ell=0}\rho(\ell)e^{-j\frac{2\pi\ell d}{N}}\frac{1-e^{j\frac{2\pi\ell d}{N}}}{1-e^{j\frac{2\pi d}{N}}}\nonumber \\
&=&\frac{-1}{N(1-e^{j\frac{2\pi d}{N}})}\bigg(\sum^{L-1}_{\ell=0}\rho(\ell)e^{-j\frac{2\pi\ell d}{N}} -\sum^{L-1}_{\ell=0}\rho(\ell)\bigg)\nonumber \\
&=&\frac{1-\bar{\rho}(d)}{N(1-e^{j\frac{2\pi d}{N}})},
\eea
where the vector $\bar{\brho}=[\bar{\rho}(0),\ldots,\bar{\rho}(N-1)]^{\rm T}$ contains the $N$-point DFT samples of the channel power delay profile. Therefore, the asymptotic ICI power $P_{\rm ICI}(d)$ of (\ref{eqn:PICI_MRC}) can be obtained as
\be\label{eqn:PICI_MRC2}
P_{\rm ICI}(d)=\frac{|1-\bar{\rho}(d)|^2}{4N^2\sin^2(\pi d/N)}.
\ee

\subsection{ISI power calculation}\label{subsec:PISI_MRC}
From (\ref{eqn:H_ISI}), one may realize that the off-diagonal elements of the matrices $\widetilde{\bH}_{{\rm ISI},m}$ have a phase difference of $\pi$ compared with those of the matrices $\widetilde{\bH}_{{\rm ICI},m}$. Thus, through the same line of derivations as in Subsection~\ref{subsec:PICI_MRC}, the ISI power due to the off-diagonal elements of the matrices $\widetilde{\bH}_{{\rm ISI},m}$ can be obtained as 
\be\label{eqn:PISI_MRC}
P_{\rm ISI}(d)=\frac{|1-\bar{\rho}(d)|^2}{4N^2\sin^2(\pi d/N)},~~~{\rm for}~~d \neq 0.
\ee
Using the diagonal elements of $\widetilde{\bH}_{{\rm ISI},m}$ from (\ref{eqn:H_ISI}) in a similar way as in Subsection~\ref{subsec:Ps_MRC}, for $d=0$, we get $P_{\rm ISI}(0)=(\frac{\bar{\tau}}{N})^2$.

\subsection{Asymptotic SIR derivation}\label{subsec:SIR_MRC}
Using the results of Subsections~\ref{subsec:Ps_MRC} to \ref{subsec:PISI_MRC}, as the number of BS antennas approaches infinity, the SIR at the receiver output for all the subcarriers approaches the asymptotic value 
\bea\label{eqn:SIR_MRC}
{\rm SIR}&=&\frac{P_{\rm s}}{P_{\rm ICI}+P_{\rm ISI}}\nonumber \\
&=&\frac{(1-\frac{\bar{\tau}}{N})^2}{(\frac{\bar{\tau}}{N})^2+\sum^{N-1}_{d=1}\frac{|1-\bar{\rho}(d)|^2}{2N^2\sin^2(\pi d/N)}}.
\eea
From (\ref{eqn:SIR_MRC}), one may realize that the asymptotic average SIR does not go to infinity. In other words, using the conventional MRC, channel distortions due to the multipath effect in the absence of CP do not completely average out as the number of BS antennas tends to infinity. Consequently, SIR saturates at a certain level, given in (\ref{eqn:SIR_MRC}), which depends on the channel statistics.

\section{Proposed TR-MRC with Post-Equalization}\label{sec:TRZF}
As it was mathematically shown in the previous section, when the conventional MRC is utilized,
 a residual interference remains and the SIR saturates at a certain level even for infinite number of BS antennas. This is due to the correlation between the combiner taps and the ISI and ICI components. In this section, we propose application of TR-MRC technique together with post-equalization of the combined signal in frequency domain to solve the SIR saturation problem. 

In TR-MRC, the received signal at each BS antenna is first prefiltered with the time-reversed and conjugated version of the CIR and then all the resulting signals are combined. Using (\ref{eqn:rm}), this procedure can be written as
\bea\label{eqn:rm2}
\br^{\rm TR}_{k} &=& \frac{1}{\sqrt{M}}\sum^{M-1}_{m=0}\br_m\ast \breve{\h}_{k,m} \nonumber \\
&=& \frac{1}{\sqrt{M}} \sum^{M-1}_{m=0}\sum^{K-1}_{j=0}(\x_j \ast\bg_{kj,m}+\bnu_m\ast\breve{\h}_{k,m})\nonumber \\ 
&=& \sum^{K-1}_{j=0} \x_j \ast \bg_{kj} + \bnu^{\rm TR}_{k},
\eea
where $\breve{\h}_{k,m}=[h_{k,m}^*(L-1),\ldots,h_{k,m}^*(0)]^{\rm T}$ is the time-reversed and conjugated version of the CIR between the $k^{\rm th}$ MT and BS antenna $m$. $ \bg_{kj}= \frac{1}{\sqrt{M}}\sum^{M-1}_{m=0}\bg_{kj,m} =  \frac{1}{\sqrt{M}}\sum^{M-1}_{m=0}\h_{j,m}\ast \breve{\h}_{k,m}=[g_{kj}(-L+1),\ldots,g_{kj}(L-1)]^{\rm T}$ is the equivalent time-reversal channel response of user $k$ when $j=k$ and the cross-talk channel response between the users $k$ and $j$ when $j\neq k$. Due to the fact that the CIRs of different users are statistically independent with respect to each other, the cross-talk channel responses tend to zero as $M$ tends to infinity and the equivalent time-reversal channel becomes an impulse at the position of the $L^{\rm th}$ sample. Finally, $\bnu^{\rm TR}_{k}=\frac{1}{\sqrt{M}} \sum^{M-1}_{m=0}\bnu_m\ast\breve{\h}_{k,m}$ is AWGN at the output of the time-reversal filter of user $k$.

With perfect timing and frequency synchronization, the time-reversal filter output for user $k$ at the position of the $i^{\rm th}$ symbol within a rectangular window of size $N$ samples and a delay of $L$ samples can be represented as
\be\label{eqn:rmi1}
\br^{\rm TR}_{k,i} = \hspace{-0.1cm}\sum^{K-1}_{j=0}\hspace{-0.15cm} \big(\bG_{kj}^{{\rm ISI}_1}\x_{j,i-1}+\bG_{kj}^{\rm ICI}\x_{j,i}+\bG_{kj}^{{\rm ISI}_2}\x_{j,i+1}\big)+\bnu^{\rm TR}_{k,i}, 
\ee
where $\bG_{kj}^{{\rm ISI}_1}$, $\bG_{kj}^{\rm ICI}$ and $\bG_{kj}^{{\rm ISI}_2}$ are $N\times N$ Toeplitz matrices comprising the ISI components due to the tail of the symbol $i-1$, the ICI components within the symbol $i$ and the ISI components originating from the beginning of the symbol $i+1$ from user $j$ to user $k$, respectively. The matrices $\bG_{kj}^{{\rm ISI}_1}$ have the same structures as that of $\bH_{0,m}$ where the values $h_m(\ell)$ are replaced with $g_{kj}(\ell)$ having the same indices, $\ell$. The $pq$ entries of the ICI matrices, $\bG_{kj}^{\rm ICI}$, can be represented as 
\be\label{eqn:G1m}
[\bG_{kj}^{\rm ICI}]_{pq}=
\left\{
	\begin{array}{ll}
	g_{kj}(p-q), &|p-q| \leqslant L-1, \\
	0, & \rm{otherwise},
	\end{array}
\right. 
\ee
and $\bG_{kj}^{{\rm ISI}_2}$ has the following structure
{\small\begin{align}
 & \bG_{kj}^{{\rm ISI}_2} = \nonumber \\
 &{\scriptsize \begin{pmatrix}
0 & 0   & \cdots & 0 &\cdots & 0 \\
\vdots &\vdots & \ddots & \vdots & \ddots & \vdots \\
0  & 0  & \cdots & 0 & \cdots & 0 \\
g_{kj}(-L+1) & 0 & \cdots & 0 & \cdots & 0 \\
g_{kj}(-L+2) & g_{kj}(-L+1) & \cdots & 0 & \cdots & 0 \\
\vdots  & \vdots  & \ddots & \vdots & \ddots & \vdots \\
g_{kj}(-1) & g_{kj}(-2) & \cdots & g_{kj}(-L+1) & \cdots & 0
 \end{pmatrix}}.
\end{align}}
After passing the signal obtained in (\ref{eqn:rmi1}) through an OFDM demodulator, we have
\begin{align}\label{eqn:dhatTR}
\hat{\dd}_{k,i}^{\rm TR} &= 
 \sum_{j=0}^{K-1} \Big(\bF_N\bG_{kj}^{{\rm ISI}_1}\bF_N^{\rm H}\bd_{j,i-1} +  \bF_N\bG_{kj}^{\rm ICI}\bF_N^{\rm H}\bd_{j,i} \nonumber \\
&+~\bF_N\bG_{kj}^{{\rm ISI}_2}\bF_N^{\rm H}\bd_{j,i+1}\Big)+\bF_N\bnu^{\rm TR}_{k,i} \nonumber \\
&= \sum_{j=0}^{K-1} \widetilde{\bG}_{kj}^{{\rm ISI}_1} \bd_{j,i-1} + \widetilde{\bG}_{kj}^{\rm ICI} \bd_{j,i} + \widetilde{\bG}_{kj}^{{\rm ISI}_2} \bd_{j,i+1} +\bar{\bnu}_{k,i}^{\rm TR},
\end{align}
where $\hat{\dd}_{k,i}^{\rm TR}=[\hat{d}_{k,i}^{\rm TR}(0),\ldots,\hat{d}_{k,i}^{\rm TR}(N-1)]^{\rm T}$ is the time-reversal estimate of the $k^{\rm th}$ user data vector over the $i^{\rm th}$ OFDM symbol and $\bar{\bnu}_{m,i}^{\rm TR}=\bF_N\bnu^{\rm TR}_{k,i}$. $\widetilde{\bG}_{kj}^{{\rm ISI}_1} = \bF_N\bG_{kj}^{{\rm ISI}_1}\bF_N^{\rm H}$, $\widetilde{\bG}_{kj}^{\rm ICI} = \bF_N\bG_{kj}^{\rm ICI}\bF_N^{\rm H}$, and $\widetilde{\bG}_{kj}^{{\rm ISI}_2} = \bF_N\bG_{kj}^{{\rm ISI}_2}\bF_N^{\rm H}$ are $N\times N$ frequency domain ISI and ICI matrices whose elements can be shown as
\bse\label{eqn:G} \begin{align}
&\left[\widetilde{\bG}_{kj}^{{\rm ISI}_1}\right]_{pq} = \bff_p^\mathrm{T} \bG_{kj}^{{\rm ISI}_1} \bff_q^* = \ba_{pq}^\mathrm{H} \bg_{kj}, \\
&\left[\widetilde{\bG}_{kj}^{{\rm ICI}}\right]_{pq}   = \bff_p^\mathrm{T} \bG_{kj}^{{\rm ICI}} \bff_q^* = \bb_{pq}^\mathrm{H} \bg_{kj}, \\
&\left[\widetilde{\bG}_{kj}^{{\rm ISI}_2}\right]_{pq} = \bff_p^\mathrm{T} \bG_{kj}^{{\rm ISI}_2} \bff_q^* = \bc_{pq}^\mathrm{H} \bg_{kj},
\end{align} \ese
where $\bff_p$ is the $p^{\rm th}$ column of the $N$-point DFT matrix and the vectors $\ba_{pq}$, $\bb_{pq}$, and $\bc_{pq}$ are determined by
\bse\label{eqn:abc} \begin{align}
\ba_{pq} = & \frac{1}{{N}}[\bzero_{1\times L},\omega^{-q},\ldots,\omega^{-q(L-1)}\sum_{i=0}^{L-2}\omega^{-(p-q)i}]^{\rm T},\\
\bb_{pq} = &\frac{1}{{N}}[\omega^{p(L-1)}\sum_{i=L-1}^{N-1}\omega^{-(p-q)i},\ldots,\sum_{i=0}^{N-1}\omega^{-(p-q)i}, \nonumber\\
&\omega^{-p}\sum_{i=0}^{N-2}\omega^{-(p-q)i},\ldots,\omega^{-p(L-1)}\sum_{i=0}^{N-L}\omega^{-(p-q)i}]^{\rm T}, \\
\bc_{pq} = & \frac{1}{{N}}[\omega^{q(L-1)}\sum_{i=N-L+1}^{N-1}\omega^{-(p-q)i},\omega^{q(L-2)}\times \nonumber\\
&\sum_{i=N-L+2}^{N-1}\omega^{-(p-q)i},\ldots,\omega^{q}\omega^{-(p-q)(N-1)},\bzero_{1\times L}]^{\rm T},
\end{align} \ese
respectively, where $\omega\triangleq e^{-j\frac{2\pi}{N}}$. Looking deeper into the interference matrices above, one may realize that their diagonal elements dominate the off-diagonal ones. This is due to the fact that the amplitude of the elements in the vectors $\ba_{pq}$, $\bb_{pq}$, and $\bc_{pq}$ have their maximum value when $p=q$ and are equal to $|\ba_{pp}| = \frac{1}{N}[\bzero_{1\times L},1,\ldots,L-1]^{\rm T}$, $|\bb_{pp}| =\frac{1}{N}[N-L,\ldots,N,N-1,\ldots,N-L]^{\rm T}$, and $|\bc_{pp}| =\frac{1}{N}[L-1,\ldots,1,\bzero_{1\times L}]^{\rm T}$, respectively. Moreover, the elements of $\ba_{pq}$, $\bb_{pq}$, and $\bc_{pq}$, implicitly show that the ICI power originating from matrices, $\widetilde{\bG}_{kj}^{{\rm ICI}}$, is much larger than the ISI power especially for large values of $N$ relative to $L$ which is usually the case in practical systems.

\begin{figure*}
\centering
\psfrag{O}[][]{{\scriptsize OFDM}}
\psfrag{M}[][]{{\scriptsize ~Modulator~}}
\psfrag{DM}[][]{{\scriptsize Demodulator}}
\psfrag{FD}[][]{{\scriptsize Frequency-Domain}}
\psfrag{and}[][]{{/}}
\psfrag{P}[][]{{\scriptsize Post-processing}}
\psfrag{ZF}[][]{{\scriptsize ZF-based}}
\psfrag{Mu}[][]{{\scriptsize Multiuser detection}}
\psfrag{Num}[][]{{\scriptsize One demodulator}}
\psfrag{OFDMdemod}[][]{{\scriptsize per user\hspace{0.4cm}}}
\psfrag{BS}[][]{{\small Base station}}
\psfrag{Tx}[][]{{\small Mobile terminals}}
\psfrag{WC}[][]{{\small Wireless channels}}
\psfrag{d1}[][]{{$\scriptstyle \dd_{0}$}}
\psfrag{dK}[][]{{$\scriptstyle \dd_{K-1}$}}
\psfrag{d1hat}[][]{{$\scriptstyle \hat{\dd}_{0}$}}
\psfrag{dKhat}[][]{{$\scriptstyle \hat{\dd}_{K-1}$}}
\psfrag{h11}[][]{{$\scriptstyle \bh_{0,0}$}}
\psfrag{hK1}[][]{{$\scriptstyle \bh_{K-1,0}$}}
\psfrag{h1M}[][]{{$\scriptstyle \bh_{0,M-1}$}}
\psfrag{hKM}[][]{{$\scriptstyle \bh_{K-1,M-1}$}}
\psfrag{v1}[][]{{$\scriptstyle \bnu_0$}}
\psfrag{vM}[][]{{$\scriptstyle \bnu_{M-1}$}}
\psfrag{TR}[][]{{\scriptsize TR-MRC}}
\psfrag{h11TR}[][]{{$\scriptstyle \breve{\bh}_{0,0}$}}
\psfrag{hK1TR}[][]{{$\scriptstyle \breve{\bh}_{K-1,0}$}}
\psfrag{h1MTR}[][]{{$\scriptstyle \breve{\bh}_{0,M-1}$}}
\psfrag{hKMTR}[][]{{$\scriptstyle \breve{\bh}_{K-1,M-1}$}}
\includegraphics[scale=0.2]{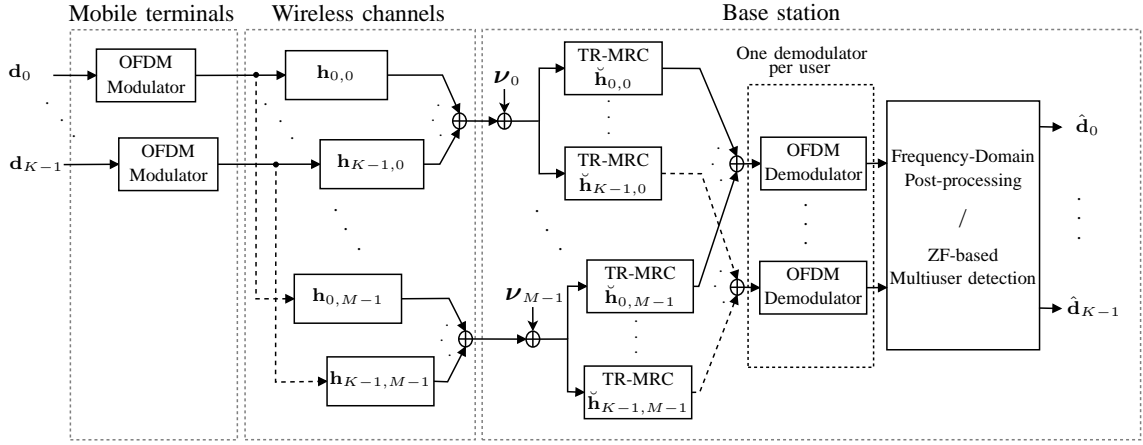}
\caption{Baseband system implementation in the uplink.}
\label{fig:Implementation}
\vspace{-0.4cm}
\end{figure*}

As mentioned earlier, the equivalent time-reversal channel becomes an impulse as $M$ tends to infinity. Thus, the ISI and ICI terms tend to zero. Moreover, MUI, i.e. the cross-talk between different users, fades away by increasing the number of base station antennas. However, in practice, the number of BS antennas is finite and as will be shown in Section~\ref{sec:Numerical_Results}, TR-MRC suffers from a large amount of MUI. Therefore, to tackle the cross-talk interference, we propose a frequency domain post-equalization based on zero-forcing criterion. 

Based on the above discussion on the ISI and ICI matrices, in equation (\ref{eqn:dhatTR}), the interference due to the off-diagonal elements of the matrices $\widetilde{\bG}_{kj}^{\rm ICI}$ and the ISI terms $\sum_{j=0}^{K-1} \widetilde{\bG}_{kj}^{{\rm ISI}_1} \bd_{j,i-1}$ and $\sum_{j=0}^{K-1} \widetilde{\bG}_{kj}^{{\rm ISI}_2} \bd_{j,i+1}$ can be absorbed into $\bar{\bnu}_{k,i}^{\rm TR}$ as additional noise. Thus, (\ref{eqn:dhatTR}) can be rearranged as
\be\label{eqn:dhatTRapprox}
\hat{\dd}_{k,i}^{\rm TR} = \sum_{j=0}^{K-1} {\BG}_{kj}^{\rm ICI} \bd_{j,i} +{\bnu'}_{k,i},
\ee
where ${\BG}_{kj}^{\rm ICI}$ is an $N\times N$ diagonal matrix with diagonal elements $[\widetilde{\bG}_{kj}^{\rm ICI}]_{pp},~p\in\{0,\ldots,N-1\}$ and ${\bnu'}_{k,i} = [\nu'_{k,i}(0),\ldots,\nu'_{k,i}(N-1)]^{\rm T}$. Using (\ref{eqn:dhatTRapprox}), for a given subcarrier $p$, we can form a linear system of equations based on the signals of all the users. Hence, we have
\be\label{eqn:LSE}
\hat{\bd}_{i}^{\rm TR}(p)= \widetilde{\BG}_p \bd_{i}(p) + \bnu'_{i}(p),
\ee
where $[\widetilde{\BG}_p]_{kj} = [\widetilde{\bG}_{kj}^{\rm ICI}]_{pp}$, $\forall k,j\in\{1,\ldots,K-1\}$, $\hat{\bd}_{i}^{\rm TR}(p) = [\hat{d}_{0,i}^{\rm TR}(p),\ldots,\hat{d}_{K-1,i}^{\rm TR}(p)]^{\rm T}$, ${\bd}_{i}(p)= [d_{0,i}(p),\ldots,d_{K-1,i}(p)]^{\rm T}$, and $\bnu'_{i}(p) = [\nu'_{0,i}(p),\ldots,\nu'_{k-1,i}(p)]^{\rm T}$. Therefore, to separate the transmitted data symbols of different users and get rid of the cross-talk components, we propose the following ZF-based post-equalization techniue for a given subcarrier $p$.
\be\label{eqn:TRZF}
\hat{\bd}_{i}^{\rm TR-ZF}(p)= \widetilde{\BG}_p^{-1}\hat{\bd}_{i}^{\rm TR}(p)=\bd_{i}(p) + \widetilde{\BG}_p^{-1}\bnu'_{i}(p),
\ee
where $\hat{\bd}_{i}^{\rm TR-ZF}(p) = [\hat{d}_{0,i}^{\rm TR-ZF}(p),\ldots,\hat{d}_{K-1,i}^{\rm TR-ZF}(p)]^{\rm T}$. As it will be shown in Section~\ref{sec:Numerical_Results}, TR-MRC accompanied with our proposed post-equalization significantly improves the SIR performance compared with utilization of TR-MRC alone. It is worth mentioning that in (\ref{eqn:TRZF}), ZF is applied considering the data symbols on the same subcarrier, $p$, within the same time-slot, $i$. This is because the interference due to the ISI and ICI is treated as noise.

Fig.~\ref{fig:Implementation}, depicts baseband implementation of our proposed TR-ZF algorithm in this paper. As one may realize, opposed to the conventional frequency domain equalization or TR-MRC implementations in the literature, \cite{Dubois2013,Maaz2015}, where the BS has one OFDM demodulator per antenna, our approach only uses one OFDM demodulator per user. This obviously simplifies the BS structure as the number of users in massive MIMO is much smaller than the number of BS antennas. Note that no CP or guard band is appended/discarded in the OFDM modulators/demodulators. As the channel estimates in the uplink are utilized for downlink transmission, our proposed TR-ZF can be applied for downlink transmission but in reverse order. Thus, the following steps are taken for downlink transmission, 1) ZF-precoding and 2) OFDM modulation, 3) TR-MRT.

\section{Numerical Results}\label{sec:Numerical_Results}
In this section, we evaluate the performance of OFDM without CP in massive MIMO for the conventional MRC and our proposed ZF-TR technique in terms of both SINR and BER. We verify the validity of our analytical derivations in Section~\ref{sec:SIR_Analysis} through simulations. In our simulations, we consider the total number of $N=256$ subcarriers, and multipath channels with the normalized exponentially decaying PDP of $\rho(\ell) = {e^{-\alpha \ell}}/\left({\sum_{i=0}^{L-1} e^{-\alpha i}}\right)$, where $L = 15$ and $\alpha = 0.1$. 

\begin{figure}[!t]
\centering
\includegraphics[scale=0.56]{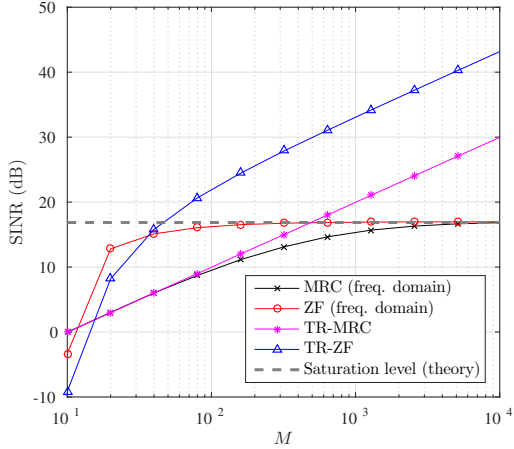}
\vspace{-0.2cm}
\caption{SINR performance comparison of different equalization methods considered in this paper when $K=10$ users are considered.}
\label{fig:sinr_K10}
\vspace{-0.1cm}
\end{figure}

Fig. \ref{fig:sinr_K10} compares the SINR performance of different equalization techniques discussed in this paper as a function of the number of BS antennas $M$ where $K=10$ active MTs are considered. We assume perfect power control for all the users where the signal to noise ratio (SNR) at the BS input is $10$ dB.
As expected, in the absence of CP, SINR saturates for the conventional MRC technique and it matches the asymptotic SIR derived in Section~\ref{sec:SIR_Analysis}. Due to the presence of MUI, the SINR of MRC saturates when $M$ becomes very large. To get rid of MUI, the conventional frequency domain ZF combining can be utilized,\cite{Rusek2013}. Since, for a very large number of BS antennas frequency domain MRC and ZF are equivalent, ZF saturates at the same level as that of MRC. This saturation happens for ZF very fast as it removes MUI. Opposed to frequency domain ZF and MRC, TR-MRC can solve the saturation problem. As one may realize from Fig. \ref{fig:sinr_K10}, while the SINR of the TR-MRC does not saturate, for the values of $M$ lower than $500$, it has a poorer performance than ZF. On the contrary, the proposed TR-ZF technique has a close SINR performance to the frequency domain ZF for the values of $M$ up to $50$. As the number of BS antennas become larger the SINR has a linear increase. The superior performance of the proposed TR-ZF technique compared to the TR-MRC is because of the frequency-domain post-processing stage where the cross-talk between the users is removed. In the present case, more than 12 dB SINR gain is achieved by using the proposed TR-ZF method. As the number of BS antennas reaches $250$, the proposed TR-ZF gains an SINR of $10$ dB higher than saturation level of the conventional MRC.

\begin{figure}[!t]
\centering
\hspace{-0.3cm}
\includegraphics[scale=0.56]{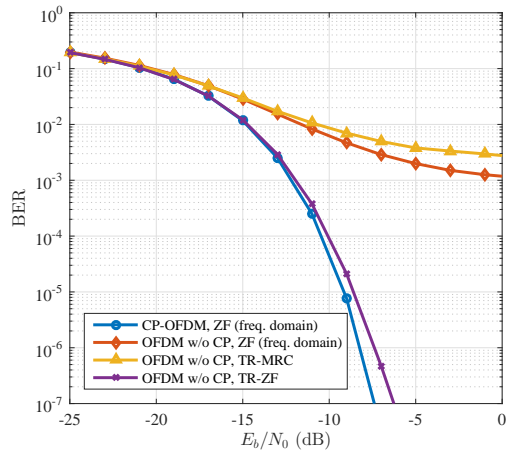}
\caption{BER performance comparison of the conventional ZF and proposed TR-ZF techniques with CP-OFDM for $K=5$ and $M=200$.}
\label{fig:ber}
\vspace{-0.5cm}
\end{figure}

In Fig. \ref{fig:ber}, we compare the BER of the conventional ZF and proposed TR-ZF techniques with that of CP-OFDM with ZF. We set CP-OFDM as a benchmark for our comparison as its SIR is infinity, thanks to the presence of a CP longer than the channel delay spread. In our BER simulations, we consider uncoded 16-QAM modulation, $K=5$ users and $M=200$ BS antennas. Based on our results, TR-ZF has a BER performance very close to CP-OFDM while having an improved bandwidth efficiency due to the absence of CP. 
The proposed TR-ZF technique has at most $0.6$ dB performance loss compared with CP-OFDM. However, the conventional ZF has a poor performance.

\section{Conclusion}\label{sec:Conclusions}
We studied OFDM without CP in massive MIMO channels to investigate if the channel distortions average out in the large antenna regime. We analyzed the SIR performance of the conventional MRC in the frequency domain as the number of base station antennas tends to infinity. From this analysis, we found that using the conventional MRC, there always exists some residual interference even for an infinite number of BS antennas. Therefore, the asymptotic SIR saturates at a certain level. To solve the SIR saturation issue, we proposed a TR technique with a frequency domain multiuser detection based on ZF criterion. Our simulations show that the SIR of our proposed TR-ZF technique is a linear function of the number of BS antennas. In addition, we numerically evaluated and compared the BER performance of the conventional MRC and the proposed TR-ZF technique with our benchmark, i.e., OFDM with CP leading to the SIR of infinity. Based on our simulations, the conventional MRC has a poor BER performance. In contrast, the BER curve of the proposed TR-ZF technique is very close to that of OFDM with less than $1$ dB performance loss at high SNRs.

\bibliographystyle{IEEEtran} 

\begin{thebibliography}{10}
\providecommand{\url}[1]{#1}
\csname url@samestyle\endcsname
\providecommand{\newblock}{\relax}
\providecommand{\bibinfo}[2]{#2}
\providecommand{\BIBentrySTDinterwordspacing}{\spaceskip=0pt\relax}
\providecommand{\BIBentryALTinterwordstretchfactor}{4}
\providecommand{\BIBentryALTinterwordspacing}{\spaceskip=\fontdimen2\font plus
\BIBentryALTinterwordstretchfactor\fontdimen3\font minus
  \fontdimen4\font\relax}
\providecommand{\BIBforeignlanguage}[2]{{%
\expandafter\ifx\csname l@#1\endcsname\relax
\typeout{** WARNING: IEEEtran.bst: No hyphenation pattern has been}%
\typeout{** loaded for the language `#1'. Using the pattern for}%
\typeout{** the default language instead.}%
\else
\language=\csname l@#1\endcsname
\fi
#2}}
\providecommand{\BIBdecl}{\relax}
\BIBdecl

\bibitem{Marzetta2010}
T.~Marzetta, ``{Noncooperative cellular wireless with unlimited numbers of base
  station antennas},'' \emph{IEEE Transactions on Wireless Communications},
  vol.~9, no.~11, pp. 3590--3600, 2010.

\bibitem{Rusek2013}
F.~Rusek, D.~Persson, B.~K. Lau, E.~Larsson, T.~Marzetta, O.~Edfors, and
  F.~Tufvesson, ``{Scaling up MIMO: Opportunities and challenges with very
  large arrays},'' \emph{IEEE Signal Processing Magazine}, vol.~30, no.~1, pp.
  40--60, Jan 2013.

\bibitem{Larsson2014}
E.~Larsson, O.~Edfors, F.~Tufvesson, and T.~Marzetta, ``Massive {MIMO} for next
  generation wireless systems,'' \emph{IEEE Communications Magazine}, vol.~52,
  no.~2, pp. 186--195, 2014.

\bibitem{Bjornson2015}
E.~Bjornson, M.~Matthaiou, and M.~Debbah, ``Massive {MIMO} with non-ideal
  arbitrary arrays: {Hardware} scaling laws and circuit-aware design,''
  \emph{IEEE Transactions on Wireless Communications}, vol.~14, no.~8, pp.
  4353--4368, 2015.

\bibitem{Xiang2015}
X.~Gao, O.~Edfors, F.~Tufvesson, and E.~Larsson, ``Massive {MIMO} in real
  propagation environments: {Do} all antennas contribute equally?'' \emph{IEEE
  Transactions on Communications}, vol.~63, no.~11, pp. 3917--3928, Nov 2015.

\bibitem{Molisch2007}
A.~Molisch, M.~Toeltsch, and S.~Vermani, ``Iterative methods for cancellation
  of intercarrier interference in {OFDM} systems,'' \emph{IEEE Transactions on
  Vehicular Technology}, vol.~56, no.~4, pp. 2158--2167, 2007.

\bibitem{Beheshti2009}
M.~Beheshti, M.~Omidi, and A.~Doost-Hoseini, ``{Equalisation of SIMO-OFDM
  systems with insufficient cyclic prefix in doubly selective channels},''
  \emph{IET Communications}, vol.~3, no.~12, pp. 1870--1882, December 2009.

\bibitem{Gomes2008}
J.~Gomes, A.~Silva, and S.~Jesus, ``{OFDM} demodulation in underwater
  time-reversed shortened channels,'' in \emph{IEEE OCEANS}, Sept 2008, pp.
  1--8.

\bibitem{Zhiqiang2012}
Z.~Liu and T.~C. Yang, ``On the design of cyclic prefix length for
  time-reversed {OFDM},'' \emph{IEEE Transactions on Wireless Communications},
  vol.~11, no.~10, pp. 3723--3733, October 2012.

\bibitem{Dubois2013}
T.~Dubois, M.~Helard, M.~Crussiere, and I.~Maaz, ``Time reversal applied to
  large {MISO-OFDM} systems,'' in \emph{International Symposium on Personal
  Indoor and Mobile Radio Communications (PIMRC)}, Sept 2013, pp. 896--901.

\bibitem{Maaz2015}
M.~Maaz, M.~Helard, P.~Mary, and M.~Liu, ``Performance analysis of
  time-reversal based precoding schemes in {MISO-OFDM} systems,'' in \emph{IEEE
  Vehicular Technology Conference (VTC Spring)}, May 2015, pp. 1--6.

\bibitem{Pitarokoilis2015}
A.~Pitarokoilis, S.~Mohammed, and E.~Larsson, ``Uplink performance of
  time-reversal {MRC} in massive {MIMO} systems subject to phase noise,''
  \emph{IEEE Transactions on Wireless Communications}, vol.~14, no.~2, pp.
  711--723, 2015.

\bibitem{Pitarokoilis2012}
------, ``On the optimality of single-carrier transmission in large-scale
  antenna systems,'' \emph{IEEE Wireless Communications Letters}, vol.~1,
  no.~4, pp. 276--279, 2012.

\bibitem{Farhang2014}
A.~Farhang, N.~Marchetti, L.~Doyle, and B.~Farhang-Boroujeny, ``Filter bank
  multicarrier for massive {MIMO},'' in \emph{IEEE Vehicular Technology
  Conference (VTC Fall)}, Sept 2014, pp. 1--7.

\end{thebibliography}

\end{document}